\documentclass[%
 reprint,
 amsmath,amssymb,
 aps,
]{revtex4-2}

\usepackage{booktabs}
\usepackage[x11names]{xcolor}
\usepackage{graphicx}
\usepackage{dcolumn}
\usepackage{bm}
\usepackage{svg}
\usepackage{appendix}
\usepackage[hidelinks]{hyperref}

\begin{document}

\newcommand{\amend}{\color{Firebrick2}}

\preprint{APS/123-QED}

\title{Parameter Inference in Non-linear Dynamical Systems via Recurrence Plots and Convolutional Neural Networks}

\author{L. Lober}
    \email{luiza.lober@usp.br}
    \affiliation{%
     Departamento de Matemática Aplicada e Estatística, Instituto de Ciências Matemáticas e de Computação, Universidade de São Paulo—Campus de São Carlos, Caixa Postal 668, 13560-970 São Carlos, São Paulo, Brazil
    }%

\author{M. S. Palmero}
    \email{palmero@usp.br}
    \affiliation{%
     Departamento de Matemática Aplicada e Estatística, Instituto de Ciências Matemáticas e de Computação, Universidade de São Paulo—Campus de São Carlos, Caixa Postal 668, 13560-970 São Carlos, São Paulo, Brazil
    }%
    
\author{F. A. Rodrigues}
    \affiliation{%
     Departamento de Matemática Aplicada e Estatística, Instituto de Ciências Matemáticas e de Computação, Universidade de São Paulo—Campus de São Carlos, Caixa Postal 668, 13560-970 São Carlos, São Paulo, Brazil
    }%

\date{\today}

\begin{abstract}
     Inferring control parameters in non-linear dynamical systems is an important task in analysing general dynamical behaviours, particularly in the presence of inherently deterministic chaos. Traditional approaches often rely on system-specific models and involve heavily parametrised formulations, which can limit their general applicability. In this study, we present a methodology that employs recurrence plots as structured representations of non-linear trajectories, which are then used to train convolutional neural networks to infer the values of the control parameter associated with the analysed trajectories. We focus on two representative non-linear systems, namely the logistic map and the standard map, and show that our approach enables accurate estimation of the parameters governing their dynamics. When compared to regression models trained directly on raw time-series data, the use of recurrence plots yields significantly more robust results. Although the methodology does not aim to predict future states explicitly, we argue that accurate parameter inference, when combined with predetermined initial conditions, enables the reconstruction of a system’s evolution due to its deterministic nature. These findings highlight the potential of recurrence-based learning frameworks for the automated identification and characterisation of non-linear dynamical behaviours.

    \begin{description}
    \item[Keyworks]
        Chaos; Neural Network; Recurrence Plots; logistic map; standard map.
    \end{description}
\end{abstract}

\maketitle


\section{\label{sec:intro} Introduction}

Identifying and characterising the diverse dynamical behaviours of non-linear dynamical systems is a challenge, particularly due to the inherent complexity arising from chaotic dynamics \cite{Strogatz2000, Stewart2000}. Even in the deterministic scenario, these systems can exhibit multiple distinct dynamics, including periodic, quasi-periodic and chaotic behaviours \cite{Zaslavsky1992}. Periodic behaviours repeat themselves regularly, whereas quasi-periodic dynamics display structured but non-repeating patterns \cite{Lichtenberg1992, Ott2002}. Chaotic dynamics, on the other hand, evolve unpredictably over time due to their extreme sensitivity to initial conditions (ICs), despite being governed by fully deterministic rules \cite{Zaslavsky1985, Wiggins2003}. Understanding these different behaviours and accurately inferring their governing parameters is essential for gaining deeper insight into the fundamental properties of such systems.

One way to approach such problem is to benefit from the significant developments to neural networks architectures that have been made in the last decade \cite{Shrestha2019Apr}, with the introduction of several robust and versatile architectures that allows one to solve high-dimensional partial differential equations (PDEs) \cite{Lawal2022Nov}, identify jets in high-energy physics \cite{Khoda2023Apr} and more broadly for the accurate identification of images in classification problems \cite{Taye2023Mar}. Recently, the relevance of neural networks has been highlighted by the 2024 Nobel Prize in Physics, awarded to Geoffrey E. Hinton and John J. Hopfield. Convolutional Neural Networks (CNN) \cite{Cong2023Mar, Li2021Jun} are a subset of feed-forward neural networks that are widely used in image, text and audio recognition tasks, providing a strong foundation for natural language processing techniques that are now state-of-the-art in artificial intelligence, while remaining relevant due to their effectiveness in many areas. In non-linear dynamical systems, CNNs have also been shown to identify chaotic dynamics when combined with recurrence analysis, achieving over 92\% accuracy \cite{Zhou2023Aug, Nam2020Dec}.

Parallel to these advancements, Recurrence Plots (RPs) have become an established approach for visualising and analysing the recurrence of states within dynamical systems \cite{Webber2005, Marwan2007, Marwan2023}. Different dynamical regimes generate distinct visual recurrence patterns, allowing for the qualitative identification of behaviours such as periodic, chaotic, or even stochastic dynamics \cite{Donner2011, Webber2015}. Because RPs are effectively binary images, with pixels representing recurrent or non-recurrent states, they are particularly suitable as inputs for CNNs. Although RPs have been widely employed in the detection and analysis of chaotic dynamics \cite{Goswami2019, Prado2020, Pham2020, Palmero2022}, their potential for inferring control parameters solely and directly from the recurrence patterns remains relatively unexplored. 

Several approaches have been applied to parameter identification and classification tasks in dynamical systems. Traditional techniques such as Kalman filtering \cite{Kalman1960, Grewal2011} have been used in linear and weakly non-linear contexts, providing robust state estimation and parameter tracking capabilities. Similarly, Non-linear Auto-regressive Moving Average with eXogenous inputs (NARMAX) models \cite{Billings2013, Chen1989} have proven effective for capturing complex non-linear behaviours, enabling simultaneous model identification and parameter estimation from observed data. A comprehensive overview of more recent developments is provided by Quaranta et al. \cite{Quaranta2020}, reviewing state-of-the-art methods for parameter inference in complex systems. Among these, deep learning techniques that include Long Short-Term Memory (LSTM) networks and reservoir computing \cite{Pathak2018, Vlachas2020} have emerged as powerful tools for both forecasting and parameter inference in non-linear dynamical systems. In addition, Estebsari et al. \cite{Estebsari2020} implement a NN framework with inputs similar to those we propose, while Kumar and Kostina \cite{Kumar2025} introduce a Huber loss-based parameter estimation scheme that significantly outperforms classical methods for the Lorenz system. Within this context, our approach differentiates itself by employing RPs as alternative feature spaces for characterising dynamical regimes. Rather than relying solely on raw temporal sequences, we explore how image-based representations of dynamical trajectories enable efficient, data-driven identification of control parameters using CNN-based architectures, also facilitating the analysis of higher-dimensional systems through the RP representation.

Our study introduces a novel approach that combines RPs with CNNs to infer control parameters governing non-linear dynamical behaviours. We apply it to two well-known dynamic systems, the logistic map and the standard map, chosen for their representative non-linear characteristics. By training CNNs on fitting RPs generated from these maps, we assess how accurately the underlying control parameters can be estimated from visual recurrence patterns alone. Additionally, we evaluate the ability of our approach to effectively classify different dynamical regimes. Due to the deterministic nature of these systems, we argue that accurate inference of control parameters, combined with predetermined initial conditions, provides sufficient information to reconstruct their dynamical evolution, even though the method itself does not explicitly predict future states of these systems. 

The remainder of this paper is organized as follows. Section\ \ref{sec:method_models} provides detailed descriptions of the methodology, including the construction of RPs, the dynamical characteristics of the logistic and standard maps, and the CNN architecture. Section\ \ref{sec:results} presents and discusses our results, analysing the performance of CNNs in the control parameter inference and regime classification tasks. Appendices\ \ref{appendix_A} and\ \ref{appendix_B} include additional discussions concerning comparisons with raw time-series regression and considerations for optimal parameter selection. Finally, Section\ \ref{sec:conclusions} summarizes the key findings of this work and suggests possible directions for future research, including potential extensions to more complex and realistic dynamical systems.

\section{\label{sec:method_models} Methodology and Models}

This section is divided into three sub-sections. Firstly, the process of constructing the RPs is outlined, providing a detailed explanation of how a given percentage of recurrences is determined. Next, the two dynamical systems of interest, namely the logistic and standard maps, are described in detail. Finally, a discussion of the chosen neural network architecture and metrics is presented. Essentially, the whole methodology has been designed to recover the characteristic control parameters of the models through efficient pattern recognition of the RPs.

\subsection{\label{subsec:rp} Recurrence Plots}

\begin{figure*}
    \includegraphics[scale=0.45]{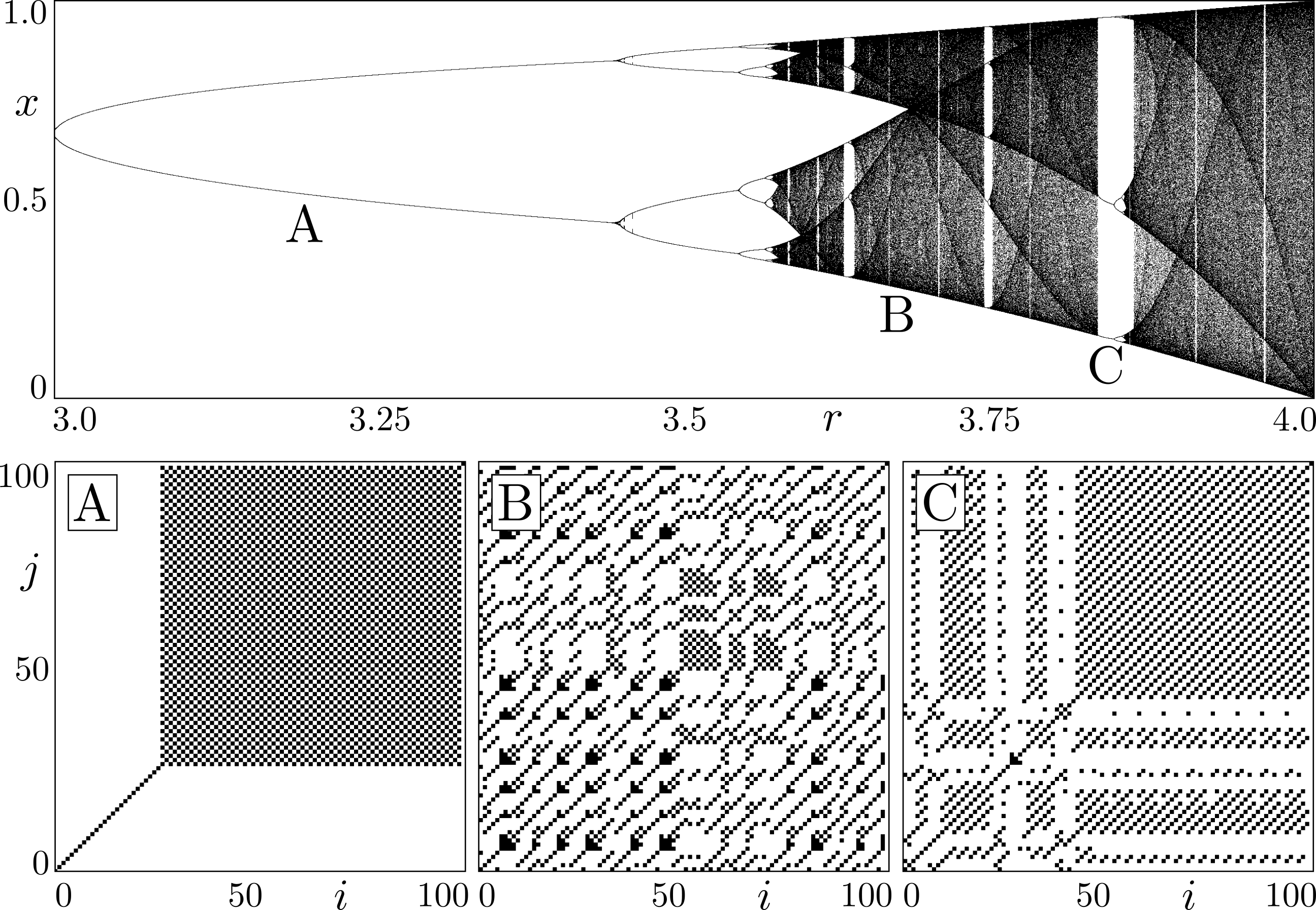}
    \caption{(Upper panel) Typical bifurcation diagram of the logistic map; (Lower panels) Three RPs from different values of the control parameter $r$. For A, the pair IC and parameter was set $(x_0 = 0.1, r = 3.20)$ -- permanent periodic dynamics; In B, the pair was $(x_0 = 0.1, r = 3.70)$ -- chaotic dynamics; In C, $(x_0 = 0.1, r = 3.85)$ -- chaotic, but close to stability islands. RPs were build from $T=100$ and fixed $RR=20\%$. 
    \label{fig:logistic_rp}} 
\end{figure*}

The essence of the proposed methodology relies on the generation of multiple distinct Recurrence Plots (RPs), which serve as fundamental visual representations of the various non-linear dynamical behaviours of interest. A recurrence is typically defined as the return of a trajectory \(\boldsymbol{x}(t_j)\) to a neighbourhood of a previously visited state \(\boldsymbol{x}(t_i)\), where \(t_i < t_j\). In other words, a state at time \(t_j\) is considered recurrent if it approximately revisits a past state. Given that this analysis is performed over a finite-time window, we emphasize that \(0 < t_i < t_j \leq T\), where \(T\) denotes the maximum iteration time of the trajectory.

In the case of one-dimensional dynamical systems, such as the logistic map, constructing an RP requires transforming the scalar time series into a multi-dimensional representation. This is accomplished via \textit{time-delay embedding}, a technique inspired by Takens' embedding theorem \cite{Takens1981, Noakes1991}. The process involves reconstructing the system's state space by introducing delayed copies of the original time series. Given a scalar sequence $u(t)$, the embedded representation is defined as 

\begin{equation}
    \mathbf{U}(t) = \left[ u(t), u(t + \tau), u(t + 2\tau), \dots, u(t + (m-1)\tau) \right],
\end{equation}where $\tau$ is the time delay, and $m$ is the embedding dimension. For $\tau = 1$ and $m=2$, each state in the embedded space consists of the value $u(t)$ and its immediate successor $u(t+1)$, capturing direct temporal relationships between consecutive states, which is sufficient for this methodology. Once the time series is embedded in this higher-dimensional space, an RP is constructed by identifying recurrent states based on a predefined threshold distance, forming the Recurrence Matrix (RM).

For two-dimensional systems, such as the standard map, this additional embedding step is unnecessary. Since the system's phase space is inherently well-defined—consisting of the generalized pair position-momentum, the RM can be directly extracted by computing distances between trajectory points. 

Depending on the model, its control parameters, and the chosen initial conditions, distinct dynamical behaviours may emerge. To ensure a fair and meaningful comparison across chaotic, periodic, and quasi-periodic dynamics, all RPs were constructed using a fixed recurrence rate ($RR$), the percentage of recurrent points in the plot \cite{Marwan2007}. This normalization allows for an unbiased analysis, ensuring that differences in recurrence structures arise from the underlying dynamics rather than arbitrary choices of threshold distances.

The process of building RPs with the same recurrence percentage begins with computing a pairwise distance matrix $D \in \mathbb{R}^{T \times T}$, where each element $D_{ij}$ defined as follows
    \begin{equation}
        D_{ij} = \|\varphi_i - \varphi_j\|,
        \label{eq:distance_matrix}
    \end{equation}
\noindent
where $\varphi_i = \boldsymbol{\varphi}(t_i)$, $\varphi_j = \boldsymbol{\varphi}(t_j)$ and $\|\cdot\|$ is a suitable norm. Typically, the Euclidean norm is often used for measuring distances in the respective phase space, which was also employed for this methodology. Note that the $i$ and $j$ are time indexes.

To ensure that the RP has a desired $RR$, a suitable distance threshold $\varepsilon$ must be determined, such that the proportion of points within this distance is equal to the $RR$. Mathematically, this requirement can be expressed as
    \begin{equation}
    \frac{1}{N^2} \sum_{i=1}^N \sum_{j=1}^N \Theta(\varepsilon - D_{ij}) = RR,
        \label{eq:requirement_percentage}
    \end{equation}
\noindent
where $\Theta(\varepsilon - D_{ij})$ is the Heaviside step function, which is equal to 1 if $(\varepsilon - D_{ij}) \geq 0$, and 0 otherwise. 

The next step to determine $\varepsilon$ is to consider the set of all distinct pairwise distances from the distance matrix $D$, excluding the diagonal elements since $D_{ii} = 0$ by definition. Let the sorted set of these distances be denoted as
    \begin{equation}
    \{d_1, d_2, \dots, d_M\}, ~\text{where~} M = \frac{N(N-1)}{2}.
        \label{eq:sortting_D}
    \end{equation}

The threshold $\varepsilon$ is then selected such that approximately $RR \times M$ of the distances are less than or equal to $\varepsilon$. Specifically, for a rate $RR = p$, the particular threshold $\varepsilon = \varepsilon_p$ is chosen such that it corresponds to the $\lfloor p \times M \rfloor$-th smallest value in the sorted set of distances as follows 
\begin{equation}
    \varepsilon = \varepsilon_p = d_{\lfloor p \times M \rfloor}.
    \label{eq:setting_eps}
\end{equation}

Once the best suited $\varepsilon$ is determined, the aforementioned binary Recurrence Matrix, composed by the elements $R_{i,j}$, $R \in \{0, 1\}^{T \times T}$, can be computed in terms of the distance matrix $D$ as follows 
    \begin{equation}
    R_{ij} = \Theta(\varepsilon_p - D_{ij}).
        \label{eq:RM}
    \end{equation}
\noindent
Notice that RM is also a symmetric matrix, as the pairwise distances $D_{ij} = D_{ji}$ for most commonly used norms, ensuring that recurrence relationships are bidirectional. Additionally, while selecting $\varepsilon = \varepsilon_p$, the recurrence rate of this RM is ensured to be $RR = p$.  

Finally, the visual representation of the RM is the RP of interest. Typically, RPs represent each zero entry ($R_{ij} = 0$) of the RM as a white pixel, while each $1$ entry ($R_{ij} = 1$) is represented by a black pixel. To ensure that each recurrent point is represented by a unique pixel, the desirable resolution of an RP is $T \times T$ pixels, where $T$ is the maximum iteration time of the dynamics being analysed. This resolution ensures that the recurrence patterns are accurately represented without overlap, making it easier to visually distinguish between different dynamical behaviours. The optimal choice of resolutions and $T$ is further explored in the Appendix \ref{appendix_B}.  

\subsection{\label{subsec:models} Dynamical systems}

The two dynamical systems studied in this paper are well known non-linear maps. The first, introduced by Lorenz in the 1960s \cite{Lorenz1964} and popularized by May in 1976 \cite{May1976}, the logistic map, is a classic example of how complex and chaotic behaviour can arise from a simple one-dimensional non-linear equation, with applications ranging from population dynamics, pseudo-random number generators to bifurcation theory~\cite{Strogatz2000}.

In addition to the logistic map, the Chirikov standard map, an example of a two-dimensional near-integrable Hamiltonian system, was also selected. First conceptualized in the 1970s in the context of plasma physics \cite{Chirikov1969, Chirikov1979}, the map exhibits a rich variety of dynamical behaviours, in particular Hamiltonian chaos, making it an ideal subject for this investigation.

The next subsections provide a detailed explanation along with examples of the dynamics presented by the logistic and standard maps.

\subsubsection{\label{logistic_map} The logistic map}

As briefly introduced earlier, the map brings its non-linearity from a simple quadratic term as follows

    \begin{equation}
            x_{t+1} = r x_t (1-x_t),
            \label{eq:logistic}
    \end{equation}
\noindent
where the defining control parameter $r$ in the range of [0,4] produces values of $x_t$ bounded on [0,1], with divergences presented when $r>4$. Here, it will be restricted to the former range.

The logistic map can have several behaviours according to the values taken by $r$, which are briefly described below.

\begin{itemize}
    \item $0 < r < 1.0$: Independently of the initial conditions, $\lim_{t\rightarrow \infty}{x_t} \rightarrow 0$;
        
    \item $1.0 < r < 2.0$: $\lim_{t\rightarrow \infty}{x_t} \rightarrow \frac{r-1}{r}$, also without regard to the initial conditions;
    
    \item $2.0 < r < 3.0$: Similar to the previous range, however presenting fluctuations around the final value before convergence;
        
    \item (A) $3.0 < r < 3.56995$: Permanent oscillatory behaviour is observed, with the number of modes of oscillation depending on $r$;
    
    \begin{itemize}
        \item (A1) $3.0 < r < 3.44949$: The dynamics oscillates between two values;
        
        \item (A2) $3.44949 < r < 3.54409$: Four oscillatory solutions are observed;
        
        \item (A3) $3.54409 < r < 3.56995$: Oscillations increase even further, allowing for $\kappa$ values, with $\kappa$ even and increasing according to $r$.
        
        \end{itemize}
        \item (B) $3.56995 < r < 3.82843$: onset of chaos;
        
        \item (C) $3.82843 < r < 4.0$: certain values of $r$ result in the emergence of \textit{islands of stability}.     
    \end{itemize}

For this study, data generation and analysis were restricted to values $r \in [3.0,4.0]$, and the previous ranges will be employed to define multiple labels (A1, A2, A3, B and C) for the upcoming classification task.

Figure \ref{fig:logistic_rp} shows the classical bifurcation diagram along with three distinct RPs, each considering different values of $r$ that properly display the diversity of dynamical behaviours. In agreement with the discussions presented at Sec.\ \ref{subsec:rp} and Appendix \ref{appendix_B}, all RPs from the dynamics of the logistic map were constructed considering a fixed $RR=20\%$, and $T=100$ iterations. 

    \begin{figure}[!t]
        \centering
        \includegraphics[scale=0.6]{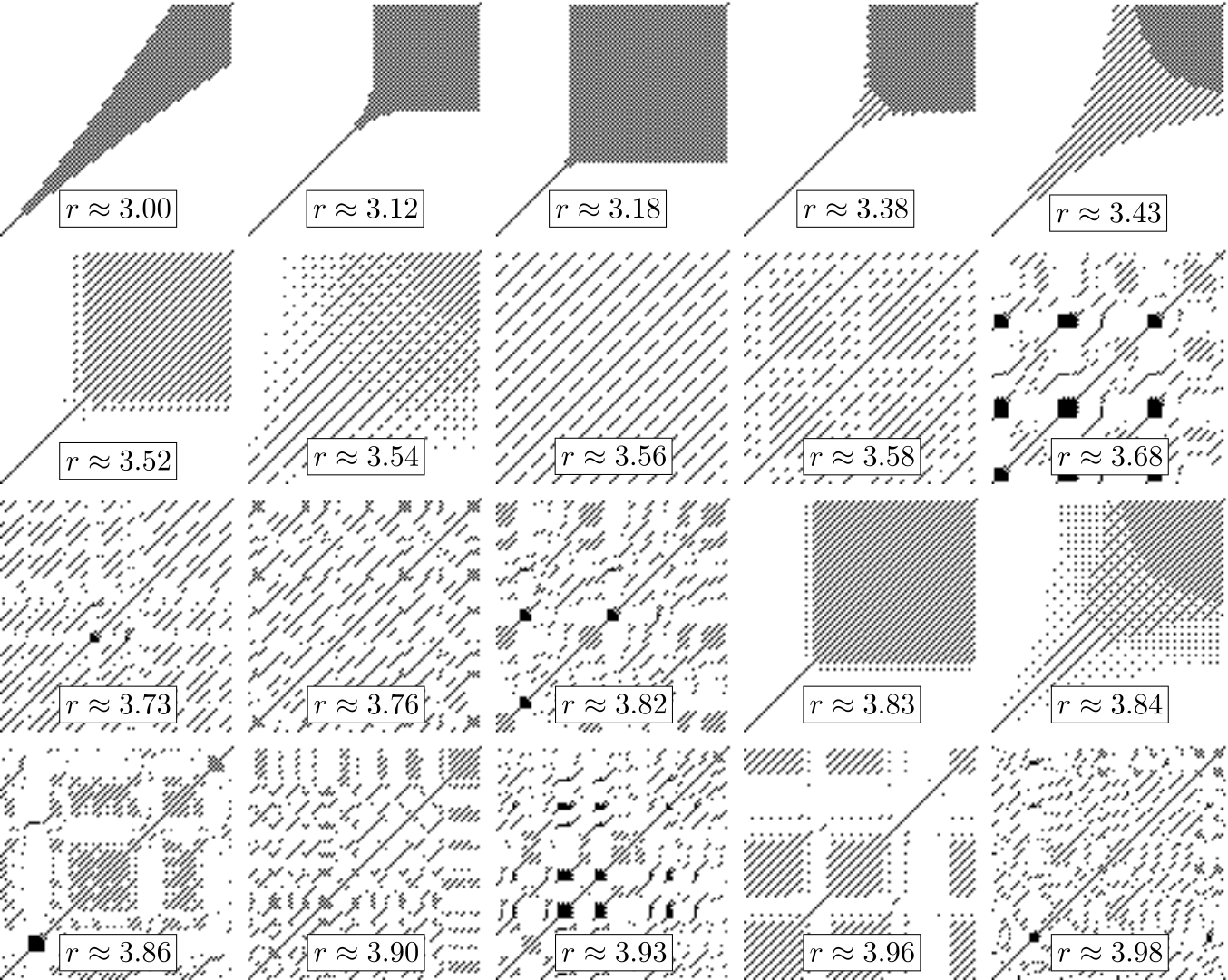}
        \caption{Twenty different dynamical behaviours displayed by their respective RP constructed from the dynamics of the logistic map. All RPs were build from $T=100$ iterations and fixed $RR=20\%$. All correspondent approximated value of the parameter $r$ is displayed inside each RP.\label{fig:logistic_examples}} 
    \end{figure}

    \begin{figure*}
        \centering
        \includegraphics[scale=0.45]{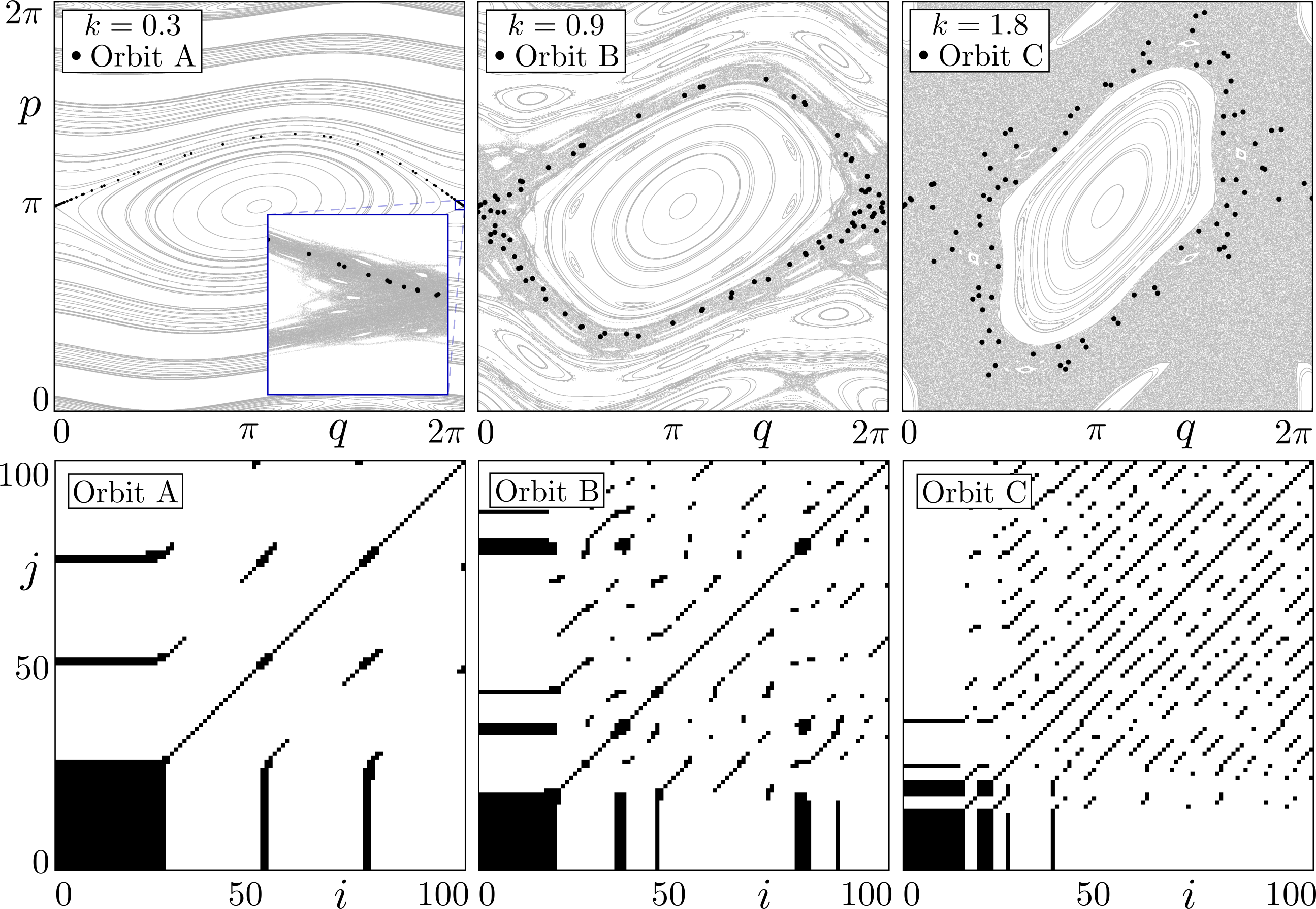}
            \caption{(Upper panels) Phase spaces of the standard map (gray background) for three increasing values of the control parameter $k$, along with a chosen chaotic orbit depicted by the black points; (Lower panels) RPs of each of these orbits. Orbits A, B and C were evolved from a fixed IC $(q_0,p_0)$ set as $(10^{-9},\pi)$, while the RPs were build from $T=100$ and fixed $RR=10\%$. 
        \label{fig:standard_rp}} 
    \end{figure*}

Since it is important to produce multiple distinct RPs, knowing that neural networks benefit greatly from large volumes of samples, a thorough numerical experiment was conducted to extract as many behaviours from the map as possible. The range $r\in[3.0,4.0]$ was uniformly divided into $10^4$ values, generating the same $10^4$ different RPs, each differing in the value of $r$ up to the fifth decimal place. The parameters of the dynamics, namely the IC $x_0$, and the maximum iteration time $T$ were set to $0.1$ and $100$ respectively. The RP parameters, including the $RR$ and resolution, were set to $20\%$ and $100\times100$ respectively. A sub-sample of this analysis, consisting of $20$ RPs, is shown in Fig.\ \ref{fig:logistic_examples}, and all $10^4$ images were compressed into a Graphic Interchange Format (GIF) made available in the supplemental material.

It is worth remarking that the initial periodic behaviours, displayed by the first row of RPs in Fig.\ \ref{fig:logistic_examples}, may exhibit patterns that do not necessarily match the typical appearance of fully periodic RPs \cite{Marwan2007, Webber2005}. This is because, instead of using a predefined threshold distance $\varepsilon$, these RPs were constructed with a fixed $RR$. Hence, as periodic dynamics presents multiple recurrences, especially for $r\approx 3.0$, the RP is proportionally filled by points until reaching the desired $RR=20\%$. It is also interesting to observe the diversity of patterns within the chaotic regime, followed by an unambiguous periodic behaviour produced by the stability islands for $r \approx 3.83$, before returning to chaos as $r$ increases.

\subsubsection{\label{standard_map} The standard map}

The standard map, also known as the Chirikov standard map, can be used to describe the motion of a particle constrained to a movement on a ring while kicked periodically by an external field. It is possible to define a sympletic non-linear discrete map that gives the particle's generalized position $q$ and momentum $p$ for the $(t+1)^{th}$ iteration by the following equations

    \begin{align}
        p_{t+1} &= p_t +k\sin(q_t) \quad \bmod(2\pi), \label{eq:stdmap_p}\\
        q_{t+1} &= q_t + p_{t+1} + \pi \quad \bmod(2\pi),\label{eq:stdmap_q}
    \end{align}
\noindent
where the parameter $k$ controls the intensity of the non-linearity and the added term $+\pi$, on the Eq.\ (\ref{eq:stdmap_q}), is to centralize the main island on the phase space. It is also important to note that this is an area-preserving map since it comes from a Hamiltonian system, and the determinant of its Jacobian matrix is equal to unity. 

The characteristic mixed phase spaces of the standard map are shown in Fig.\ \ref{fig:standard_rp}. These phase spaces are said to be mixed as chaotic areas can coexist with periodic regions \cite{Ott2002}. As illustrated by the gray points in the background, depending on the pair of ICs $(q_0,p_0)$ a given trajectory may fill invariant spanning curves i.e. curves that span throughout the whole phase space; fill periodic structures within stability islands, also known as Kolmogorov-Arnold-Moser (KAM) islands \cite{Lichtenberg1992}; or fill the chaotic area. 

As predicted by the theory of near-integrable Hamiltonian systems, chaos emerges in the vicinity of unstable fixed points as the control parameter increases \cite{Meiss1987}. It can be observed that Eqs.\ (\ref{eq:stdmap_q}) and (\ref{eq:stdmap_p}) have a stable fixed point at $(\pi, \pi)$ and an unstable one at $(0, \pi)$ or $(2\pi, \pi)$ due to the modulo $(2\pi)$ condition. Hence, for sufficiently large values of $k$, the region in the near surroundings of $(0, \pi)$ will always be filled by chaotic orbits, as evidenced by the inset zoom on the first panel of Fig.\ \ref{fig:standard_rp}, where $k= 0.3$. Figure \ref{fig:standard_rp} also displays the growth of the chaotic region for larger parameter values, namely $k=0.9$ and $k=1.8$ for the remaining upper panels.

The second panel of Fig.\ \ref{fig:standard_rp}, shows a particular phase space configuration close to an important topological transition that happens in the standard map. For the critical value $k = k_c = 0.971635$, the last invariant spanning curve is destroyed and the map transits from local to global chaos \cite{Greene1979, MacKay1983}, where for a sufficient long evolution, any chaotic orbit may fill out the whole phase space, with the only exception to the inside of the KAM islands embedded to what is now called the \textit{chaotic sea}. This configuration is exemplified by the third panel of Fig.\ \ref{fig:standard_rp}.

In addition to the mixed phase spaces, the lower panels of Fig.\ \ref{fig:standard_rp} also display three RPs associated with orbits A, B and C. These particular trajectories, represented by the black points on their respective phase spaces, are chaotic orbits initiated from a fixed IC $(q_0, p_0) = (10^{-9}, \pi)$, meaning that they start extremely close to the unstable fixed point. Indeed, at the beginning of their dynamical evolution, the influence of the fixed point is still dominant, as evidenced by the initial black squares in all RPs. Nevertheless, since it is a neighbourhood of an unstable fixed point, the chaotic behaviour eventually emerges as the orbits lose stability and explore the available chaotic areas.

\begin{figure}[!t]
        \centering
        \includegraphics[width=0.45\textwidth]{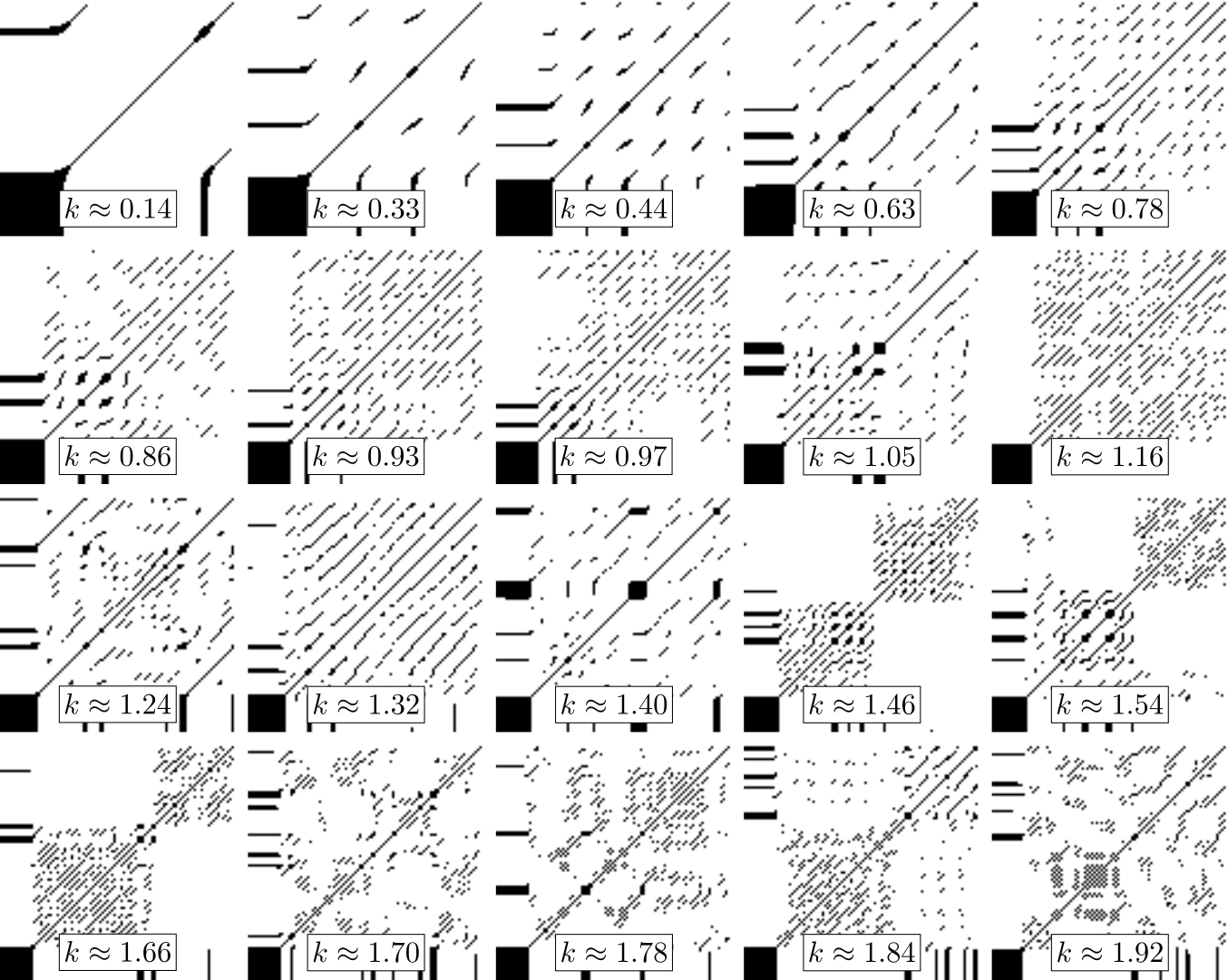}
        \caption{Twenty different chaotic behaviours displayed by their respective RP constructed from the dynamics of the standard map. All RPs were build from $T=100$ iterations and fixed $RR=10\%$. All correspondent approximated value of the parameter $k$ is displayed inside each RP.
        \label{fig:standard_examples} } 
\end{figure}

\begin{figure*}[]
    \includegraphics[scale = 0.55]{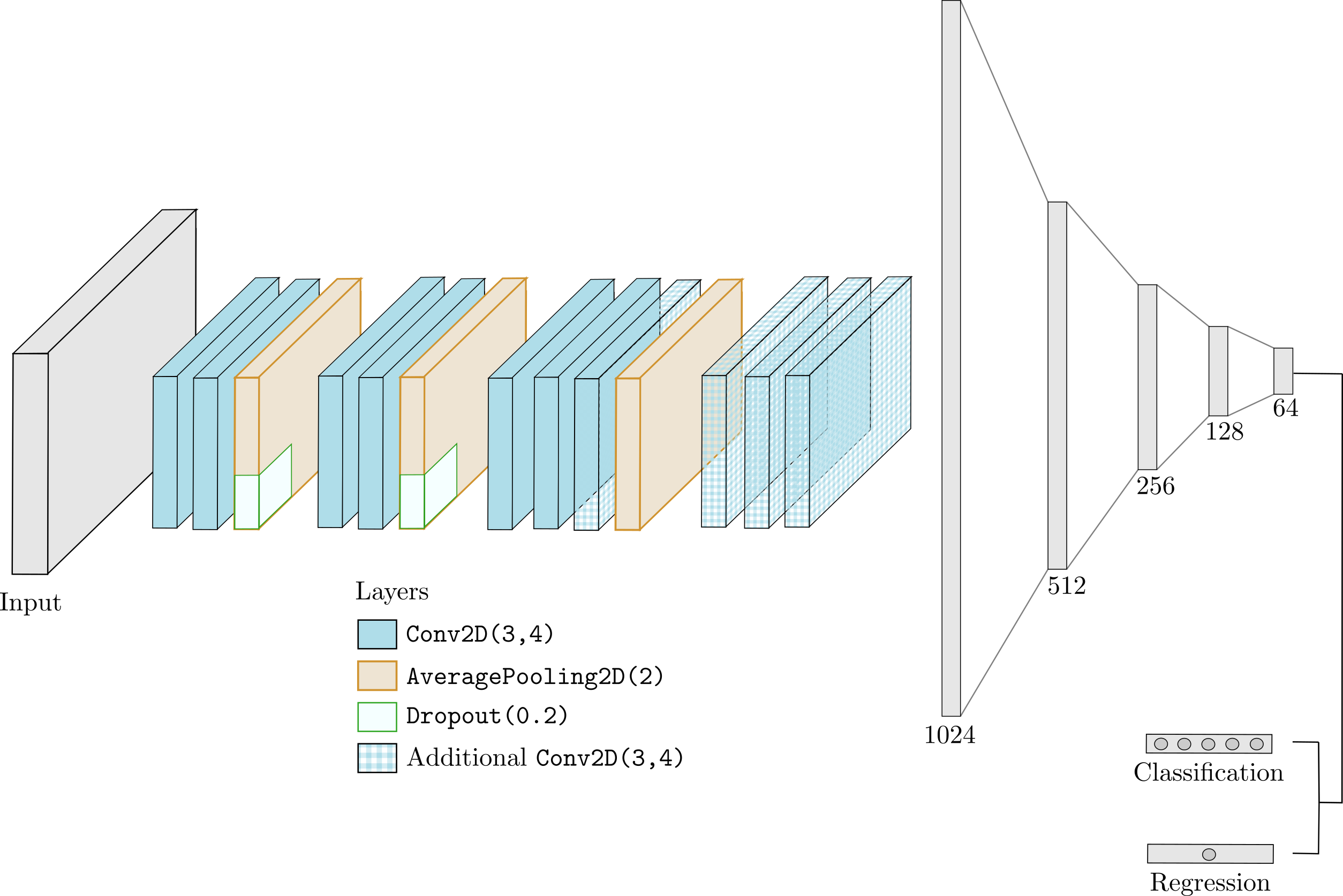}
    \caption{\label{fig:nn_architecture} General architecture of the CNN implemented. Layers indicated in solid colours were used for both dynamics, with the hatched blocks employed solely for the dynamics of the standard map, and dropout used only for the logistic map.}
\end{figure*}

Following the approach used for the logistic map in the previous subsection, Fig.\ \ref{fig:standard_examples} shows the diversity of RPs constructed from various chaotic trajectories in the standard map. This figure also employs ICs with trajectories fixed as $(q_0, p_0) = (10^{-9}, \pi)$, while the control parameter $k$ was changed thoroughly within the range $[0.1, 2.0)$. Similar to the logistic map, this range was uniformly divided into $10^4$ values, generating $10^4$ different RPs, each differing in the value of $k$ up to the fourth decimal place. The RP parameters, including the $RR$ and resolution, were set to $10\%$ and $100\times100$ respectively. While Fig.\ \ref{fig:standard_examples} shows twenty examples of these RPs with increasing values of $k$, all $10^4$ images were compressed into a GIF, which is also available in the supplementary material.

It is important to point out that these sets of RP images, exemplified by Figs.\ \ref{fig:logistic_examples} and \ref{fig:standard_examples}, provide effective visual representations of various finite-time dynamical behaviours. In this sense, understanding all the recurrence patterns associated with these RPs is essential to their characterization. To achieve this, a convolutional neural network is trained to effectively obtain the characteristic control parameters associated with these dynamical behaviours.

\subsection{\label{subsec:NN} Convolutional Neural Networks and metrics}

The architecture of all convolutional neural networks employed in this investigation was defined as shown in Fig. \ref{fig:nn_architecture}, using the structure available at TensorFlow \cite{tensorflow2015_whitepaper}. As for the choice of hyper-parameters, the model consists of: 100 batch size; \texttt{Conv2D(3,4)} for each convolutional layer; \texttt{AveragePooling2D(2)} as the pooling layers; \texttt{ReLu()} as the activation function with \texttt{Adam} as chosen optimizer. Root Mean Squared Error (RMSE) and Categorical Cross Entropy are used as loss functions for regression and classification tasks respectively. The training employed 300 maximum epochs, allowing for early stopping with the criteria of interruption being 0.001 and 20 epochs. The training set was also split in a 70\% to 30\% ratio, with the smaller subset being used for validation. The evaluation of validation loss was also used to arrive at the best combinations of the number of layers and parameters to be employed in the following simulations.

In terms of the amount of samples generated that are being used as inputs, the main models had access to $4\times 10^4$ unique RPs for the training step, which were composed of $10^4$ uniformly chosen values of $r$ and $k$ for four sets of ICs: $x_0 = 0.10, 0.25, 0.50$ ~\text{and}~$0.75$ for the logistic map; $(q_0, p_0) = (\delta \times 10^{-\gamma}, \pi)$ where $\gamma = 8$ is fixed and $\delta = 2, 4, 6 ~\text{and}~8$ for the standard map. Different values of $\gamma$ were also considered and are further explained in Appendix \ref{appendix_B}. 

As a way to quickly access the control parameters in the training stage, those are saved while naming each sample, and latter being imported to the respective train and test vectors while loading the images to the models.

Each network was trained independently in two different RP train sets: the first set containing data from the logistic map generated with parameters $r \in [3.0,4.0]$, and the second with $r$ immediately after the onset of chaos; while for the standard map, $k \in [0.1, 0.971635)$ and a larger range $k \in [0.1, 2.0]$ were used to generate the data sets for this dynamic. The amount of samples, originated from the same sets of ICs, remained constant for the analysis focused on the onset of chaos and $k<k_c \approx 0.971635$, resulting in larger training sets for these regions where the chaotic behaviour is predominant. 

When evaluating the trained neural networks, $10^3$ non-interloping samples were generated as the hold-out test set by randomly sorting the defining parameter of each map and fixing the IC $x_0=0.4$ (logistic) and $(q_0, p_0) = (5\times 10^{-8}, \pi)$ (standard).

The evaluation of the performance of the trained models used the following metrics: the weighted $f_1$ score was employed to assess the accuracy of inferences across the defined intervals of $r$ to which the testing RP samples belonged in the logistic map. For the regression tasks, a point-wise error, based on the absolute differences between the actual and inferred parameter values, was used to evaluate the estimated $r$ and $k$ values for the logistic and standard maps, respectively.

Additionally, considering an acceptable margin $m$, a given inference can be categorized as follows

\begin{equation}
    C_i(m) = \Theta(m - |y_i - \hat{y_i}|),
    \label{eq:correct}
\end{equation}where $y_i$ and $\hat{y_i}$ represent the actual and inferred parameter values for sample $i$, respectively. The Heaviside step function ensures that if $|y_i - \hat{y_i}| < m$, $C_i = 1$, indicating that the inferred value is within the acceptable margin $m$ from the actual value; otherwise, $C_i = 0$, representing an incorrect inference.

The total number of inferences within the $m$-tolerance is then given by $C(m) = \sum_{i=1}^N C_i(m)$, where $N$ is the total number of test samples. Furthermore, the percentage of correct inferences is defined as

\begin{equation}
    c(m) = \frac{C(m)}{N} \times 100\%,
    \label{eq:correct_}
\end{equation}where, in practical terms, if $m=0.01$, the percentage $c(0.01)$ expresses how accurately the NN can, indeed, infer the parameter values up to the third decimal place.

The next section is dedicated to presenting and explaining the main results of the proposed methodology.


\section{\label{sec:results} Results and discussions}

The first set of results comes from applying the aforementioned methodology to the logistic map. $4\times 10^4$ RPs, representing distinct non-linear dynamical behaviours from Eq. (\ref{eq:logistic}), were constructed considering fixed $RR = 20\%$ and $T=100$. The task of the trained NNs is to infer both the value and range of the characteristic control parameter $r$, respectively through a regression and classification task. Table \ref{tab:results_logistic} presents the outcome of the models for the hold-out test samples.

    \begin{table}[!t]
        \centering
        \caption{\label{tab:results_logistic} Results of the classification task on the logistic map, employing $f_1$ score as the metric; and the regression task, using RMSE and $c(m)$ defined in Eq.\ (\ref{eq:correct_}), both for the test set, and for each interval of $r$ as described in Sec.\ \ref{logistic_map}.}
        \begin{tabular}{@{}cccccc@{}}
        \toprule
        Class & RMSE & $c(0.1)$ & $c(0.01)$ & $f_1$ score & support  \\ \midrule
        \multicolumn{6}{c}{$3.0<r<4.0$} \\    
        \hline
        (A1)      & 0.0375   &  98.2\%  &  42.8\%  & 0.997            & 439 \\
        (A2)      & 0.0375   &  99.6\%    &  13.2\%  & 0.989            & 93 \\
        (A3)      & 0.0606   &  91.7\%    &   9.4\%  & 0.952            & 22 \\
        (B)      & 0.0235   &  100\%   &   36.6\%  & 0.983            & 265  \\ 
        (C)      & 0.0322   &  100\%   &   4.5\%  & 0.976            & 181 \\ 
        weight. avg. & 0.0416    &  -        &   -       & 0.988            & 1000 \\
        \hline
        \multicolumn{6}{c}{$3.56<r<4.0$} \\    
        \hline
    
        (A3)      & 0.0475   &  95.7\%    &   22.9\%  & 0.971            & 17      \\
        (B)      & 0.0206   &   100\%  &   37.3\%  & 0.988            & 608      \\ 
        (C)      & 0.0235   &  100\%   &   0\%  & 0.981            & 375        \\ 
        weight. avg. & 0.0334    &  -        &   -       & 0.985            & 1000 \\ \bottomrule
        \end{tabular}
    \end{table}

    \begin{figure}[!h]
        \includegraphics[scale=0.7]{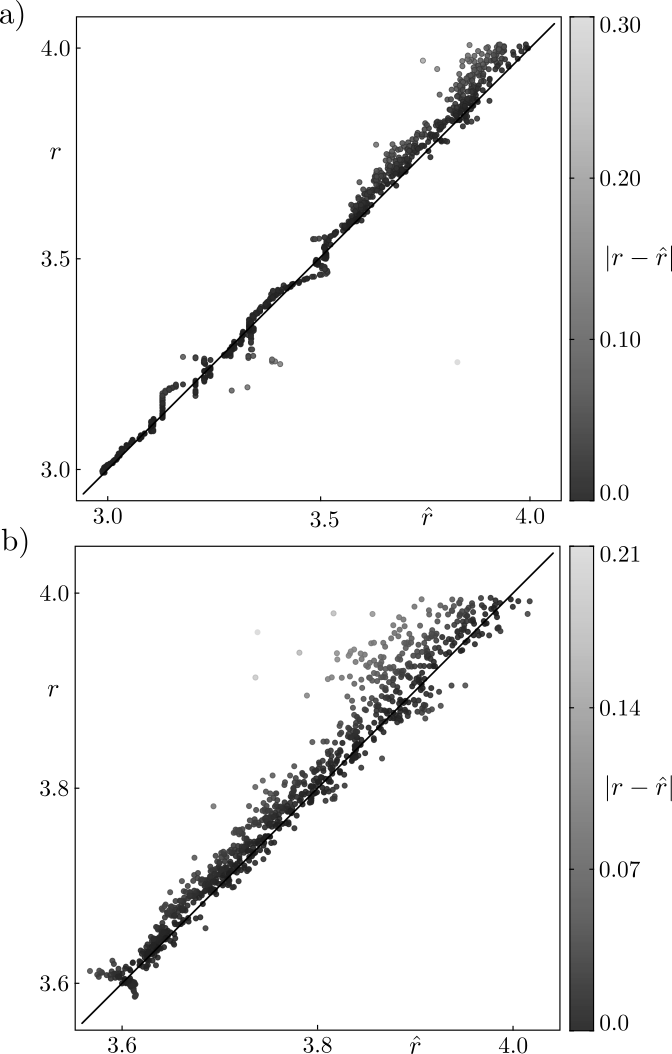}
        \caption{\label{fig:regression_logistic} Inferred values ($\hat{r}$) compared to actual parameter ($r$) values of the test set from the logistic map in the regression task. The colour axis illustrates the dispersion on the absolute difference $|r - \hat{r}|$ in both cases: (a) $r\in[3.0,4.0]$ and; (b) $r\in[3.56,4.0]$.}
    \end{figure}

Notice that the weighted accuracy obtained, shown in column ``$f_1$ score'' for training the NN almost exclusively on the chaotic region is higher than currently reported in literature \cite{Cong2023Mar, Li2021Jun}, and also improved while comparing with RPs originated from $r \in [3.0,4.0]$. As for the regression results displayed on columns ``RMSE, $c(0.1)$ and $c(0.01)$'', the same improvement on characterizing chaotic dynamics is observed when focusing solely on the chaotic region, where at least $86.93 \%$ of inferred parameters $\hat{r}$ are precise within two decimal places to the actual parameter $r$.

Figure~\ref{fig:regression_logistic} illustrates the regression results for the logistic map. In Fig.\ \ref{fig:regression_logistic}~(a), despite the presence of a distinct light-grey point far from the main regression trend, where the proposed NN inferred a value likely associated with the periodic windows within region (C), it is clear that the NN can successfully infer the parameter values, particularly before the region (B), onset of chaos at $r \approx 3.57$. Beyond this point, the inferences become less accurate as the chaotic regime dominates the dynamics. To further investigate, the same number of RP samples was used to train the NN, but considering only the range $r \in [3.56, 4.0]$. The regression results for this specific range are shown in Fig.\ \ref{fig:regression_logistic}~(b).

In comparison to (a), Fig.\ \ref{fig:regression_logistic} (b) shows an inferior agreement between the inferred and actual parameter values. This is due to the fact that the chaotic regime may present many distinct recurrence patterns, as earlier evidenced by the RPs within this range in Fig.\ \ref{fig:logistic_examples}. These many faces of chaotic behaviour are expected for a finite-time analysis, making the effective parameter inference more challenging, and consequently lowering the NN accuracy. Additionally, regions between $3.8 < r < 3.9$, where the agreement is even less accurate, correlate to the rise of periodic and quasi-periodic behaviours due to islands of stability of said region. Yet, as observed in Tab.\ \ref{tab:results_logistic}, this region is successfully classified, presenting relatively high $f_1$ score. 

It is worth mentioning that all results presented for the logistic map are further compared with those obtained from raw time-series data analysis using an equivalent NN architecture in Appendix \ref{appendix_A}. These comparisons highlight that the proposed RP-based methodology provided better parameter inferences.

    \begin{figure}[!t]
        \includegraphics[scale=0.7]{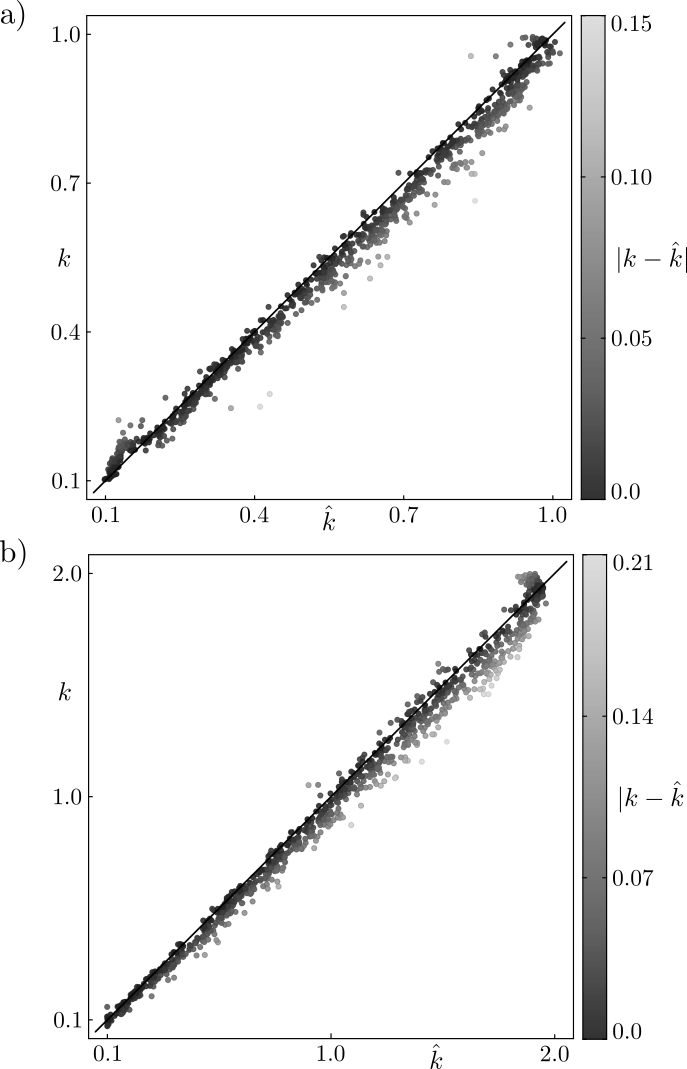}
        \caption{\label{fig:regression_standard} Inferred values ($\hat{k}$) compared to actual parameter ($k$) values of the test set from the standard map in the regression task. The colour axis illustrates the dispersion on the absolute difference $|k - \hat{k}|$ in both cases: (a) $k\in[0.1, k_c)$ and; (b) $k\in[0.1,2.0)$.}
    \end{figure}

    \begin{table}[!t]
        \centering
        \caption{\label{tab:results_standard} RMSE and overall precision $c(m)$ for the regression task for RPs generated from the standard map.}
        \begin{tabular}{@{}ccccc@{}}
        \toprule
        $k$ interval & RMSE & $c(0.1)$  &  $c(0.01)$ &  \\ \midrule
        $[0.1,k_c)$     & 0.0348   &  98.8\%  &   24.0\%   &  \\
        $[0.1,2)$      & 0.0677    &  85.5\%  &   11.8\%   &         \\ \bottomrule
        \end{tabular}
    \end{table}

Considering the standard map, Tab.\ \ref{tab:results_standard} summarizes the performance of the trained NN with the RPs from the different non-linear dynamical behaviours of Eqs.\ (\ref{eq:stdmap_p}) and (\ref{eq:stdmap_q}). Again, $4 \times 10^4$ RPs were constructed considering $T=100$, but now $RR = 10\%$. In contrast to the logistic map, all RPs of the standard map represent chaotic behaviours, specifically different finite-time trajectories in Hamiltonian chaos. In that sense, employing the process outlined in Sec.\ \ref{subsec:rp} for a relatively high $RR$ may require longer computational simulations, so $RR=10\%$ was chosen. Additionally, since only chaotic behaviour is analysed, classification tasks were neglected.

It is important to emphasize that by selecting trajectories with specific ICs, namely $(q_0, p_0) = (\delta \times 10^{-8}, \pi)$, where $\delta = 2, 4, 6 ~\text{and}~8$, is possible to ensure chaotic behaviour for $k>0.1$. As detailed in Sec.\ \ref{subsec:models}, for $k \gtrsim 0.1$, a \emph{chaotic layer} surrounds the vicinity of the unstable fixed point at $(0, \pi)$ and, as $k$ increases, the phase space is filled by the expanding chaotic sea. These chaotic trajectories are suitably represented by their respective RPs as depicted earlier by Fig.\ \ref{fig:standard_examples}.

Regression results for the standard map are shown in Fig.\ \ref{fig:regression_standard}. Initially in (a), considering the interval $k\in [0.1,k_c)$, the agreement between the inferred and actual parameter values is surprisingly notable for chaotic dynamics. A slight decrease in performance is observed in (b) for parameters beyond the critical value $k_c$. Although regions above $k>2.0$ were neglected, these results illustrate that particular finite-time chaotic behaviours, associated with the different values of $k$, can be successfully characterized via the proposed methodology. 


\section{\label{sec:conclusions} Conclusions}

The primary aim of our work was to demonstrate a recurrence-based methodology for inferring control parameters and classifying dynamical regimes in non-linear dynamical systems. By representing the system’s trajectories through recurrence plots, we provided structured visual data that could be efficiently processed by convolutional neural networks. The CNN was trained to recognise recurrence patterns associated with distinct values of the system’s control parameters and successfully distinguished between different dynamical behaviours. Although the approach does not directly predict future system states, the deterministic nature of the analysed systems implies that the accurate inference of parameters, combined with predetermined initial conditions, enables a fitting reconstruction of their dynamical evolution. This highlights the use of RPs as suitable feature spaces for parameter inference and regime classification in non-linear dynamical systems.

The chosen CNN architecture demonstrated robustness in capturing and interpreting recurrence patterns from both studied systems, achieving good accuracy in inferring control parameters for the one-dimensional logistic map, as well as for the two-dimensional standard map. Performance was enhanced when training was focused specifically on critical dynamical regimes, such as chaotic intervals in the logistic map or the stronger chaotic regime in the standard map. These results indicate that future improvements in inference accuracy could be attained through expanded training datasets, refined sampling strategies, and potentially integrating transfer learning techniques.

Further extensions of the proposed methodology could involve exploring advanced neural network architectures, such as ResNet, EfficientNet, and InceptionV3. Developing large-scale databases of RPs, analogous to the ImageNet project~\cite{imagenet_cvpr09}, would likely enhance the capability to identify and characterise distinct dynamical behaviours, enabling a fitting parameter inference across a broad spectrum of real-world dynamical systems. Moreover, incorporating attention mechanisms~\cite{Vaswani2017Jun} or similar interpretability-focused methods could improve understanding by revealing how neural networks prioritise features within the RPs, providing deeper insights into the fundamental characteristics of non-linear behaviours.

\begin{acknowledgments}
    Luiza Lober thanks the support given by São Paulo Research Foundation (FAPESP) (grants number 2022/16065-3 and 2013/07375-0). Matheus Palmero also thanks FAPESP (grant number 2023/07704-5). Francisco A. Rodrigues acknowledges CNPq (grant 308162/2023-4) and FAPESP (grants 20/09835-1 and 13/07375-0) for the financial support given for this research.
\end{acknowledgments}

\section*{Code availability}
All data and code used in this work are publicly available at \url{https://github.com/luizalober/chaotic_systems_NN}.

\appendix 
\section{Raw time-series comparison}        
\label{appendix_A}

    \begin{table}
        \centering
        \caption{\label{tab:results_ts_logistic} Results of the classification task on the time-series provided by the logistic map, employing $f_1$ score as the metric; and the regression task, using RMSE and $c(m)$, both for the test set for each interval of $r$ as described in Sec.\ \ref{logistic_map}. Refer to Tab.\ \ref{tab:results_logistic} for comparison to the RP methodology.}
        \begin{tabular}{@{}cccccc@{}}
        \toprule
        Class & RMSE & $c(0.1)$ & $c(0.01)$ & $f_1$ score & support  \\ \midrule
        \multicolumn{6}{c}{$3.0<r<4.0$} \\    
        \hline
        (A1)      & 0.1522   &  33.0\%  &  0\%  & 1.000            & 455 \\
        (A2)      & 0.0511   &  100\%    &  0\%  & 0.870            & 94 \\
        (A3)      & 0.0506   &  100\%    &   0\%  & 0.382            & 26 \\
        (B)      & 0.0600   &   94.3\%   &   5.7\%  & 0.987            & 246  \\ 
        (C)      & 0.0820   &   69.3\%   &   6.1\%  & 1.000            & 179 \\ 
        weight. avg. & 0.1138    &  -        &   -       & 0.962            & 1000 \\
        \hline
        \multicolumn{6}{c}{$3.56<r<4.0$} \\    
        \hline
    
        (A3)      & 0.0627   &  100\%    &   0\%  & 0.882            & 30      \\
        (B)      & 0.0302   &   100\%  &   21.6\%  & 0.990            & 575      \\ 
        (C)      & 0.0489   &  100\%   &   0.5\%  & 0.996            & 395        \\ 
        weight. avg. & 0.0398    &  -        &   -       & 0.989            & 1000 \\ \bottomrule
        \end{tabular}
    \end{table}

    \begin{figure}
        \includegraphics[scale=0.7]{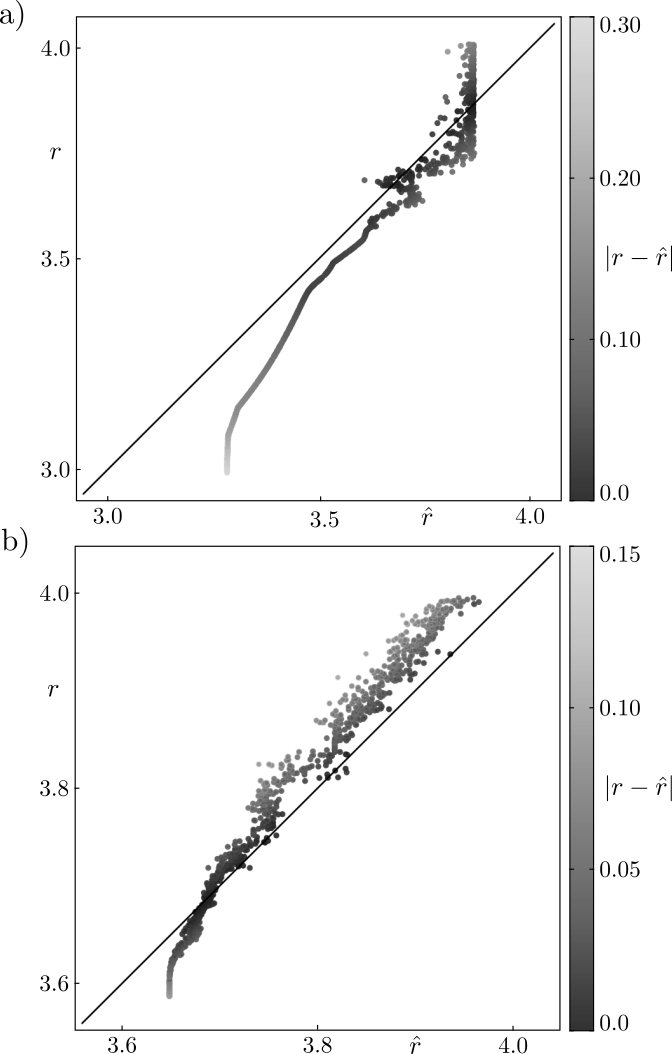}
        \caption{\label{fig:regression_series_logistic} Inferred values ($\hat{r}$) compared to actual parameter ($r$) of the test set from the logistic map in the regression task, considering raw time-series under \texttt{Conv1D(3,4)} analysis. The color axis illustrates the absolute difference $|r - \hat{r}|$. Refer to Fig.\ \ref{fig:regression_logistic} for comparison to the RP methodology.}
    \end{figure}

This appendix discusses a comparison between raw time-series analysis and the RP-based methodology outlined in Sec.\ \ref{sec:method_models}, as well as the corresponding results presented in Sec.\ \ref{sec:results}.

As mentioned at the beginning of Sec.\ \ref{subsec:rp}, the construction of RPs for one-dimensional dynamical systems, such as the logistic map, is inherently tied to their time-series representation. To ensure a suitable and meaningful comparison, we conduct an analysis using the same type of CNN but applied directly to raw time-series data. The only modifications to the network architecture presented in Fig.\ \ref{fig:nn_architecture} were made to accommodate the one-dimensional input, replacing the two-dimensional layers with \texttt{Conv1D(3,4)} and \texttt{AveragePooling1D(2)} for compatibility. The results of this time-series analysis are presented in Table \ref{tab:results_ts_logistic}.

A comparison with Table \ref{tab:results_logistic} shows that the time-series approach is outperformed in almost all $r$ classes, in both RMSE and the percentage of correct inferences $c$, particularly for control parameter values where chaos is observed. In addition, Fig.\ \ref{fig:regression_series_logistic} presents the regression results for the time-series analysis. A comparison with Fig.\ \ref{fig:regression_logistic} clearly illustrates that the RP-based methodology provides higher accuracy in inferring the values of the control parameters.

For the standard map, a direct comparison with the RP-based methodology is not as straightforward. Since it is a two-dimensional system, its natural phase space representation already incorporates the necessary information for recurrence analysis. While one could construct a time-series representation based on a derived quantity, such as the Euclidean distance between consecutive states, this would ultimately yield results closely related to those obtained via RP analysis. Therefore, while a strict analogy to the logistic map's comparison is not possible, the key advantage of RP-based analysis remains evident.

Nevertheless, the results shown in Table \ref{tab:results_ts_logistic} and Fig.\ \ref{fig:regression_series_logistic} align with previous studies \cite{Estebsari2020}, which indicate that RPs have a greater ability to encode the overall dynamical behaviour of a system compared to raw time-series analysis, further supporting our findings.

\section{Optimal parameters selection}        
\label{appendix_B}

This appendix is devoted to further explore the choice of additional parameters involved in the methodology outlined in Sec.\ \ref{sec:method_models}.

\begin{figure}
    \centering
    \includegraphics[scale=0.75]{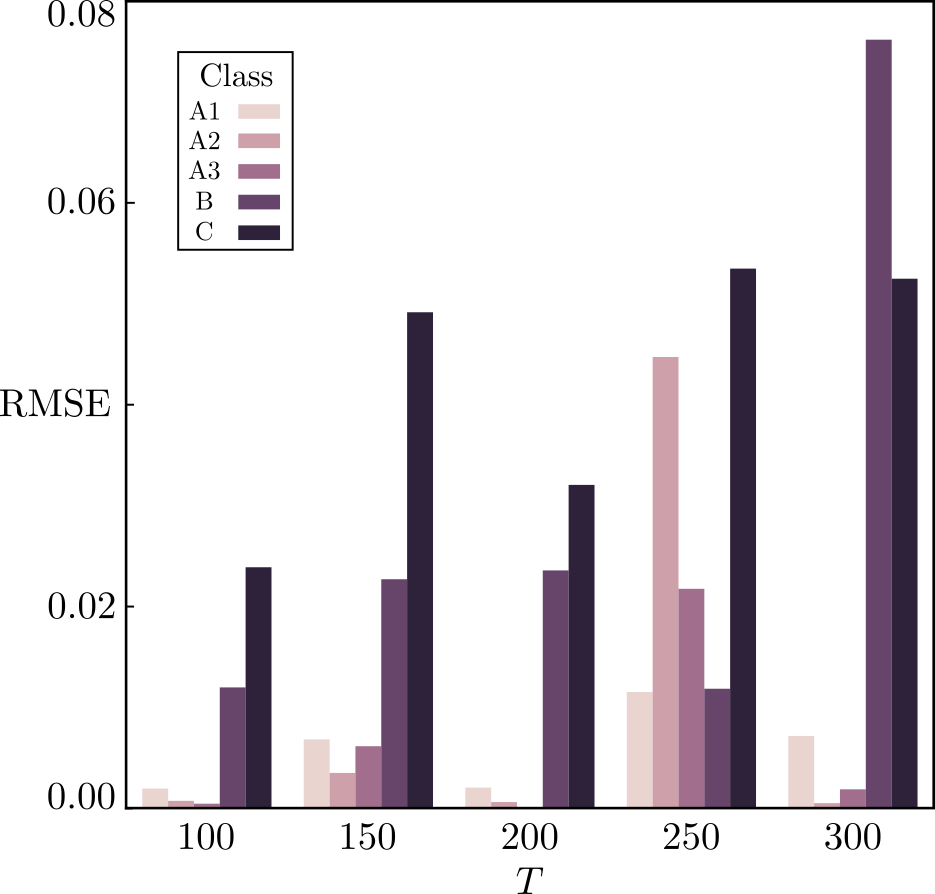}
    \caption{\label{fig:best_iter_res} Results of the Root Mean Square Error (RMSE) considering different maximum iteration number $T$ and, consequently, the RP resolution $(T \times T)$ pixels. The 5 classes are the same as presented in Sec.\ \ref{logistic_map}.}
\end{figure}

\begin{figure*}
    \centering
    \includegraphics[scale=0.75]{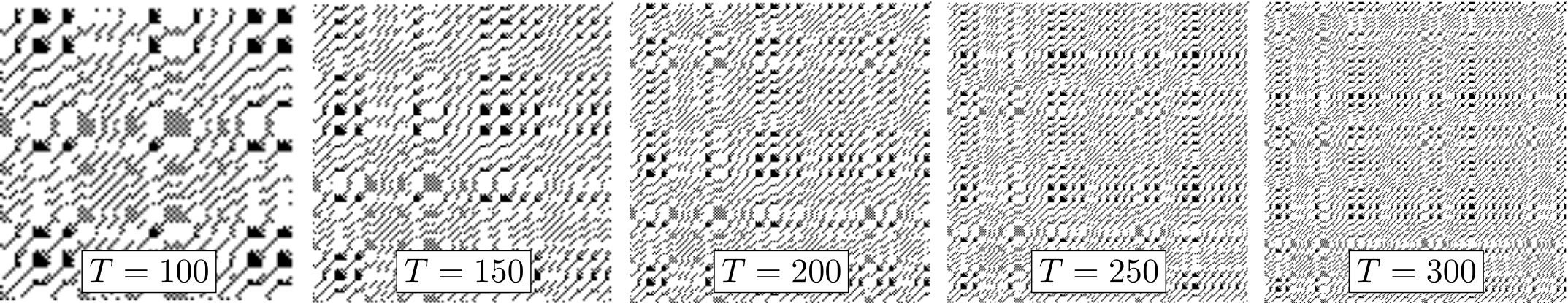}
    \caption{Five examples of RPs with increasing maximum iteration time $T$ and, consequently, their resolution ($T\times T$ pixels)\label{fig:RP_examples_T}. Here the logistic map is evolved with fixed IC $x_0 = 0.1$ and control parameter $r \approx 3.7004$.} 
\end{figure*}

\begin{figure*}
    \centering
    \includegraphics[scale=0.75]{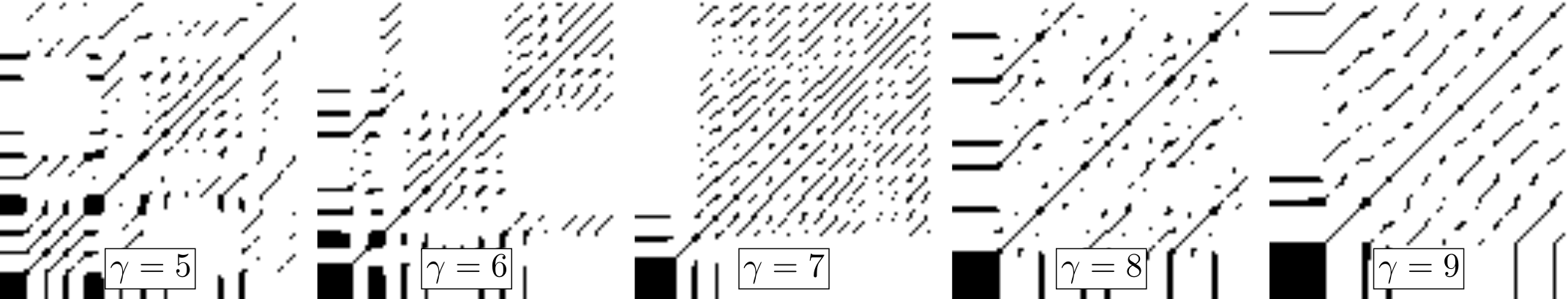}
    \caption{Five examples of RPs considering the standard map with fixed control parameter $k\approx0.97137$ and ICs $(q_0, p_0) = (\delta \times 10^{-\gamma}, \pi)$. Here $\delta = 1.0$ while varying the order $\gamma$, evolving these five trajectories up to $T=100$ iterations.  \label{fig:RP_examples_gamma} } 
\end{figure*}

The first objective is to determine the optimal pixel resolution for the RPs used by the main models. To achieve this, several numerical simulations were conducted using different sample sets of varying sizes, with their performances compared. For each comparison, every pixel represented a unique point in the RP, resulting in images with a resolution of $(T \times T)$ pixels, where $T$ denotes the number of iterations in each evolution, as defined earlier in Sec.\ \ref{subsec:rp}.

The logistic map was selected for these tests due to its ability to characterize the network's precision across various ranges, thus capturing its well-known behaviours throughout the $r$ range. The training dataset for the neural networks initially consisted of $10^4$ unique RP images, generated using uniformly selected values of $r$, while the hold-out test set contained 500 examples. These sample sizes were fixed across all simulations to manage memory consumption effectively, both in terms of RAM and VRAM, as image resolution increased. The results are presented in Fig.\ \ref{fig:best_iter_res}.

Considering the lowest values of RMSE for all classes, the optimal resolution is defined as $100\times100$ pixels, meaning also that the dynamics is evolved up to 100 iterations. This resolution not only allows for the NNs to efficiently learn the patterns of each class of the logistic map, but also capitalize on the lower computational cost of the smaller sample sizes to warrant a significant increase in the sampling size of the data, which in turn benefits the learning process of all NNs. Consequently, the sampling was enlarged to contain $4\times10^4$ unique RP images for the training set, and $10^3$ for the hold-out test set, which is the same sampling size used for the final results, as discussed in Sec.\ \ref{sec:results}.

An argument can also be made for the increased complexity of the dynamics when considering higher resolutions, which result from allowing for more iterations of the maps. These more complex RPs can compromise the efficiency of the pattern recognition capability of the NNs, which is especially true for chaotic behaviour as observed in classes B and C in Fig.\ \ref{fig:best_iter_res}. Indeed, Fig.\ \ref{fig:RP_examples_T} illustrates the compression of such patterns in an increasing pixel resolution and, consequently, maximum iteration time $T$.

    \begin{table}
        \centering
        \caption{\label{tab:results_tuning_q0} RMSE and the defined percentage of correct inferences $c(m)$ for the regression tasks considering the standard map with ICs $(q_0, p_0) = (\delta \times 10^{-\gamma}, \pi)$. Here $\delta = 1.0$ and $k\in [0.1, k_c)$.}
        \begin{tabular}{@{}ccccc@{}}
        \toprule
        $\gamma$ & RMSE & $c(0.1)$ & $c(0.01)$ &  \\ \midrule
        $5$ & 0.0365 & 98.3\% & 35.3\% &  \\
        $6$ & 0.0347 & 98.0\% & 46.9\% &  \\
        $7$ & 0.0277 & 98.9\% & 45.2\% &  \\
        $8$ & 0.0273 & 99.4\% & 47.8\% &  \\
        $9$ & 0.0410 & 98.1\% & 21.8\% &  \\ \bottomrule
        \end{tabular}
    \end{table}

The next additional parameter that needs to be carefully selected is order $\gamma$ for the distance between the pair of ICs $(q_0, p_0) = (\delta \times 10^{-\gamma}, \pi)$ and the unstable fixed point at $(0,\pi)$ in the standard map. Since only chaotic dynamics is analyzed in this case, it is fair to assume that the optimal resolution would be the same as previously chosen for the logistic map, especially considering the low RMSE for the region of chaotic behavior, namely B and C classes. With that, several simulations were performed considering the range of the characteristic control parameter $k\in [0.1, k_c)$, a fixed $\delta = 1$ and different values for the order $\gamma = 5,6,7,8 ~\text{and}~9$. The results of RMSE and the percentage of correct inferences $c(m)$ are presented in Tab.\ \ref{tab:results_tuning_q0}.

From the comparison of the results presented in Tab.\ \ref{tab:results_tuning_q0}, $\gamma = 8$ was chosen as the optimal order, meaning that the pair of ICs $(q_0,p_0)$ is effectively selected as $(10^{-8},\pi)$. As discussed in detail in Sec.\ \ref{subsec:models}, this special pair of ICs warrants chaotic dynamics even for relatively small values of the control parameter $k$, allowing a suitable comparison between different behaviours within Hamiltonian chaos. Nevertheless, the results shown in Tab.\ \ref{tab:results_tuning_q0} are expected, but also intriguing. On the one hand, RMSE values should indeed be inversely proportional to $\gamma$, meaning that the closer the ICs are to the unstable fixed point, the NN would be able to recognize even more different chaotic behaviours. On the other hand, $\gamma = 9$ produces an unexpected outcome, since it should be the lowest value of RMSE and $c(m)$. This result is still unclear and will be addressed in future works. 

Finally, Fig.\ \ref{fig:RP_examples_gamma} exemplifies the differences between the RPs of the standard map considering different orders of $\gamma$ for the pair of ICs. It is worth noting that, as expected, the size of the initial black square increases as $\gamma$ also increases, showing the influence of the unstable fixed point at the beginning of the dynamical evolution. Moreover, as an additional subject for future investigations, the rate of expansion of the black square for increasing $\gamma$ might be inherently connected to the local Lyapunov exponent associated with these chaotic trajectories.



\bibliography{bibliography}

\end{document}